\documentclass[twocolumn,showpacs,prb,superscriptaddress]{revtex4}
\newcommand{\figurewidth}{\columnwidth}
\usepackage{graphicx}
\newcommand{\ql}{q_{l}}
\newcommand{\absq}{\left|q\right|}
\newcommand{\av}{_{\mathrm{av}}}

\begin{document}

\title{Large-Scale, Low-Energy Excitations in the Two-Dimensional Ising Spin
Glass}

\author{A.~K.~Hartmann}
\email{hartmann@theorie.physik.uni-goettingen.de}
\homepage{http://www.theorie.physik.uni-goettingen.de/~hartmann}
\affiliation
{Institute for Theoretical Physics, University of G\"ottingen, 37073
G\"ottingen, Germany}
\affiliation{Department of Physics,
University of California,
Santa Cruz, California 95064}

\author{A.~P.~Young}
\email{peter@bartok.ucsc.edu}
\homepage{http://bartok.ucsc.edu/peter}
\altaffiliation{Temporary address:
Department of Theoretical Physics, 1, Keble Road, Oxford OX1 3NP,
England}
\affiliation{Department of Physics,
University of California,
Santa Cruz, California 95064}

\date{\today}

\begin{abstract}
We study large-scale, low-energy excitations in the Ising spin glass with
Gaussian interactions
in two-dimensions at zero temperature, using an optimization algorithm to
determine exact ground states. Periodic boundary conditions are applied.
Our results for the fractal
dimension of the surface, $d_s$, and stiffness exponent, $\theta'$, for
``droplet'' excitations, are
in reasonable agreement with estimates from ``domain
wall'' calculations, and so support the predictions of the ``droplet theory''.
Restricting our analysis to small lattices, we do
not find an effective value of $\theta'$ close to $-0.47$ as has been recently
proposed.
The effects of averaging over droplets of different sizes
are studied and are also found to be too small to give $\theta' \approx
-0.47$ for smaller sizes.
Larger corrections to finite-size scaling would be needed in three
and four dimensions in order for the numerical data used to support the
``TNT'' scenario to be compatible with
the droplet theory prediction that the stiffness exponent is the same for
droplets and domain walls.
\end{abstract}

\maketitle

\section{Introduction}
\label{sec:introduction}

Despite extensive studies over many years, there is still no consensus on the
nature of the spin glass state. According to the
``droplet picture''\cite{fisher:86,fisher:88,bray:86,mcmillan:84a} the lowest
energy excitation, or droplet,
of linear size $l$ containing a given site has
a characteristic energy of order $l^\theta$ where $\theta$ is a ``stiffness''
exponent.
Droplets are expected to be compact but with a surface which has a non-trivial
fractal dimension $d_s$, less than the space dimension $d$.
For Ising spins, the only case considered here, the
spin glass state persists to finite
temperature for $d \ge 3$, and in this situation one has
$\theta >0$. On the other hand
if the transition is at $T=0$, as happens in
$d = 2$, then $\theta < 0$. A common way to determine
$\theta$ is to compute the characteristic change in the ground state energy
when the boundary conditions in one direction are changed from periodic to
anti-periodic, which induces a domain wall across the system. Although the
domain wall, which stretches across the system from one side to the other, is
different geometrically from a droplet, which closes on itself, the droplet
picture makes the plausible assumption that the energy varies with length
scale in the same way for the two types of excitation. ``Domain-wall''
calculations give $\theta$ about $0.20$
in three-dimensions\cite{bray:84,mcmillan:84,hartmann:99,palassini:99},
and about $0.70$ in
four dimensions\cite{hartmann:99a,hukushima:99}. In $d=2$, a range of
domain wall
calculations\cite{mcmillan:84b,bray:84,rieger:96,palassini:99a,hartmann:01a,carter:02} give
$\theta$ about $-0.28$, see Table~\ref{tab:exponents},
which also shows
that the best estimates for
$d_s$ are around $1.27$.

\begin{table}
\begin{center}
\begin{tabular}{|l||c|c|}
\hline
Reference & $\theta$ & $d_s$ \\
\hline
\hline
Bray and Moore\cite{bray:84} &$-0.294 \pm 0.009$ & --- \\
McMillan\cite{mcmillan:84b} & $-0.281 \pm 0.005$ & --- \\
Rieger et al.\cite{rieger:96} & $-0.281 \pm 0.002$ & $1.34 \pm 0.10$ \\
Palassini and Young\cite{palassini:99a} & $-0.285\pm 0.020$ & $1.30\pm 0.08$ \\
Hartmann and Young\cite{hartmann:01a} & $-0.282 \pm 0.002$ & --- \\
Carter et al.\cite{carter:02} &$-0.282 \pm 0.003$ & --- \\
Bray and Moore\cite{bray:87} & --- & $1.26 \pm 0.03 $ \\
Middleton\cite{middleton:01} & --- & $1.27 \pm 0.01 $ \\
\hline
\end{tabular}
\end{center}
\caption{
Values of the stiffness exponent $\theta$ and surface fractal dimension $d_s$
for ``domain-wall'' calculations in two dimensions.  The estimates for
$\theta$ are consistently close to $-0.28$, which we shall take to be the
correct value (to two decimal places) for the purposes of this article. The
most accurate
estimates of $d_s$ are those of
Bray and Moore\cite{bray:87}, $1.26 \pm 0.03$, and
Middleton\cite{middleton:01}, $1.27 \pm 0.01 $.
}
\label{tab:exponents}
\end{table}

In the alternative scenario, known as ``replica symmetry breaking'' (RSB),
it is assumed that the Parisi
theory\cite{parisi:79,parisi:80,parisi:83,mezard:87} of the infinite range
Sherrington-Kirkpatrick model, applies, at least in part, to short range
systems. In this picture, the energy of droplets containing a finite
fraction of the system does not increase with increasing system size.
Although, to our knowledge, there are no calculations of the domain wall
energy within the RSB picture, there seems to be no reason why it should
vanish. Hence, it appears that different ``$\theta$-like'' exponents are
needed for domain walls and droplets within the RSB picture. To allow for this
possibility we will denote by $\theta$ the exponent for the energy of
domain walls and by $\theta'$ the exponent for droplets. Thus, in this
notation, RSB gives $\theta' = 0$.
Furthermore, in RSB, the surface of the
large-scale, low-energy excitations are expected to fill space and so have the
fractal dimension $d_s = d$. This means that the distributions of
both the ``spin overlap'' and
``link overlap'', defined in Sec.~\ref{sec:model}
below, are ``non-trivial'', in the sense that
they have a finite width in the thermodynamic limit, as opposed to the
droplet picture in which they both become ``trivial'', i.e. are just delta
functions, in this limit.

Recently, an intermediate scenario, known as TNT (for ``Trivial,
Non-Trivial''), has been proposed\cite{krzakala:00,palassini:00}, based on
numerical work in $d=3$ and 4 at zero temperature. In this picture one has
$\theta' \simeq 0 \ (< \theta)$, as in RSB, but $d_s < d$, as in the droplet
picture. Subsequently,
support for the TNT picture
was found from finite-$T$ Monte Carlo simulations\cite{katzgraber:01}.
However, it has also been
criticized by
Marinari and Parisi\cite{marinari:00} who argued that $d_s = d$ (in
which case
one obtains RSB). In addition,
Middleton\cite{middleton:01} has criticized TNT in the opposite sense by
arguing that averaging over droplets of different sizes, which was
carried out in Refs.~\onlinecite{krzakala:00}
and \onlinecite{palassini:00}, leads to an
incorrect value of $\theta$ for the small sizes studied, and that
asymptotically
$\theta' = \theta$,
so the droplet theory is correct.

Moore\cite{moore:02} has also argued that
Refs.~\onlinecite{krzakala:00} and \onlinecite{palassini:00} obtained the
result
$\theta' \simeq 0 \, (< \theta)$, rather than $\theta' = \theta$,
because
corrections to scaling were neglected. However,
the corrections that Moore considers
are different.  He argues that
there is a sub-leading (repulsive)
contribution to the energy of a droplet, varying as $l^{-\omega}$,
which arises because the wall cannot
intersect itself. The amplitude of this contribution is argued to be larger
for a droplet, since this has to return to its original location, than for a
domain wall which does not have to do this and so would have less tendency to
intersect itself. Moore argues that evidence for this contribution to the
droplet energy is found in the work of Lamarcq et al.\cite{lamarcq:02}, who
compute the energy of droplets in three dimensions by determining the
minimum energy excitation above the ground state which contains a given number
of spins including one fixed central spin. Moore further argues that a
combination of this repulsive piece of the droplet energy which decreases with
increasing size $l$, and the leading $l^\theta$ term which increases
with increasing $l$, can give a droplet energy which is roughly constant for
the small range of droplet sizes studied numerically. By contrast, for
domain walls, the amplitude of the repulsive correction
is argued to be small, and so the
correct asymptotic value of $\theta'$ (i.e. $\theta$)
would be observed even for small sizes.

Further motivation for Moore's ideas comes from results in $d=2$, where a much
larger range of system sizes $L$ can be studied than in higher
dimensions. While $T=0$
domain-wall calculations give $\theta$ about $-0.28$ as
noted above, both finite-$T$ simulations\cite{liang:92,kawashima:92a}
and $T=0$ studies
in a magnetic
field\cite{kawashima:92},
both of which excite droplets rather
than domain walls, find $\theta$ about $-0.47$. A good review of this
situation is given in Ref.~\onlinecite{kawashima:00a}. Since $\theta <0$ in
$d=2$, both the $l^\theta$ and $l^{-\omega}$
contributions to the droplet energy decrease with increasing $L$
but, Moore argues,
could combine to give an effective exponent of about $-0.47$ for
small $l$ crossing over to the asymptotic value of about $-0.28$ for larger
sizes. A different perspective on these results has in two dimensions
been given by
Kawashima\cite{kawashima:00}, who argues that $\theta' \simeq -0.47$ is the
{\em asymptotic}\/ value for droplets
and different from $\theta$ obtained from domain wall calculations.
Recently, Picco et al.\cite{picco:02} have studied the first excited state,
as well as the ground state, in a 2-$d$ spin glass, and from their results
they infer that $\theta' \simeq -0.46$.

With current algorithms\cite{hartmann:01},
it does not seem possible to push numerical studies
in $d=3$ to significantly larger sizes than has already been done.
However, in $d=2$, there are more efficient algorithms as we shall see, which
allow larger sizes to be considered, and
the goal of the present paper is to investigate
large-scale, low-energy excitations in two-dimensions at $T=0$
by the approach of Ref.~\onlinecite{palassini:00}, described below. While
Middleton\cite{middleton:01} has also applied the method of
Ref.~\onlinecite{palassini:00} to $d=2$, our work differs
from his because
we use periodic boundary conditions, as opposed to
the ``open,
with boundary spins fixed under perturbation'' or ``link periodic'' boundary
conditions used by Middleton. Since they have no surface, samples with
periodic boundary conditions are expected to have relatively small finite-size
corrections. Our choice of boundary conditions is discussed more fully in
Sec.~\ref{sec:bcs}. In addition, we also perform an analysis in which we just
look at the large droplets generated, for each lattice size, rather than
averaging over droplet sizes. 

Our motivation is to understand in detail finite-size scaling for spin glasses
in two dimensions. This should give better insights into finite-size scaling
in higher dimensions, and may help us understand
whether the result that $\theta' < \theta$ found
in three and four dimensions is real or simply caused by
corrections to scaling.
More precisely, by studying Ising spin glass ground states in $d=2$ we aim to:
\begin{enumerate}
\item[(i)]
separate out the effects of averaging over different droplet sizes, to see
how this influences the values of
$\theta$ and $d_s$ for a
system in two-dimensions with periodic boundary conditions, and
\item[(ii)]
look for the crossover, expected by Moore\cite{moore:02}, from $\theta$
about $-0.47$ to about $-0.28$.
\end{enumerate}
We consider a continuous (Gaussian) distribution of interactions to avoid the
additional complications coming from
the degenerate ground state in the
$\pm J$ model\cite{houdayer:01,hartmann:01a}.

The model and the technique to generate excitations used in
Ref.~\onlinecite{palassini:00} is described in Sec.~\ref{sec:model}. Our
choice of boundary conditions is discussed in Sec.~\ref{sec:bcs}.
The results are presented in Sec.~\ref{sec:results} and the conclusions
summarized in Sec.~\ref{sec:conclusions}.

\section{The model and technique}
\label{sec:model}

The Hamiltonian is
\begin{equation}
{\cal H} = - \sum_{\langle i,j\rangle} J_{ij} S_i S_j ,
\label{eq:ham}
\end{equation}
where the sum is over nearest neighbor pairs of sites on a square lattice,
the $S_i$ are Ising spins taking values $\pm 1$, and the $J_{ij}$ are Gaussian
variables with zero mean and standard deviation unity. The lattice contains $N
= L \times L$ sites. The techniques presented below work for several kinds
of boundary conditions, but our results will be mainly for
periodic boundary conditions.

We find the 
large-scale, low energy excitations out of the ground state
by following the procedure of Ref.~\onlinecite{palassini:00}.
First of all we compute the ground state and denote the spin configuration by
$S_i^{(0)}$.
We then add a perturbation to the Hamiltonian designed to
increase the energy of the ground state relative to the other states, and so
possibly induce a change in the ground state. This perturbation,
which depends upon a positive parameter $\epsilon$, changes the
interactions $J_{ij}$ by an amount proportional to $S_i^{(0)} S_j^{(0)}$, i.e.
\begin{equation}
\Delta {\cal H}(\epsilon)  =  \epsilon {1 \over N_b} \sum_{\langle i,j
\rangle}
S_i^{(0)} S_j^{(0)} S_i S_j,
\label{eq:deltah}
\end{equation}
where
$N_b = d N$ is the number of bonds in the Hamiltonian. In the numerical work
we took $\epsilon = 2$.

\begin{figure}
\includegraphics[width=\figurewidth]{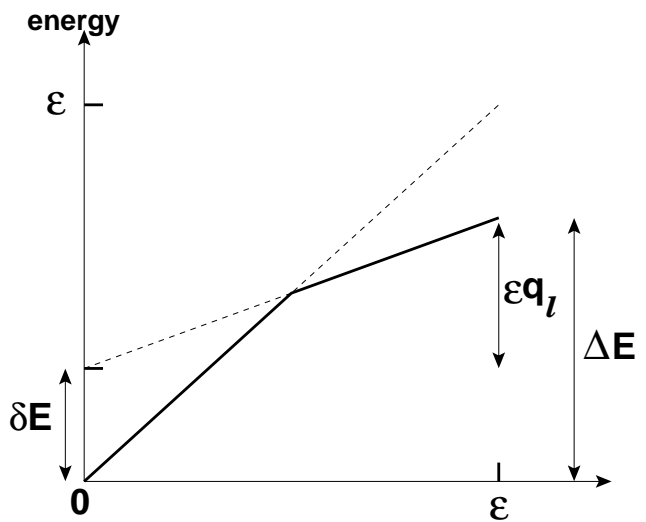}
\caption{
A sketch of the variation of the ground state and an excited state an energy
$\delta E$ above it, as a
function of the strength of the perturbation, $\epsilon$.
The variation of the ground state energy as a function of $\epsilon$
is shown by the solid line.
For the value of the perturbation indicated,
the unperturbed excited state has become the perturbed ground state.
We denote by $\Delta E$ the change in ground
state energy due to the perturbation.
}
\label{e_vs_epsilon}
\end{figure}

The energy of the ground state will thus increase exactly by an amount
$ \Delta E^{(0)} = \epsilon .$
The energy of any other state, $\alpha$ say, will increase by the lesser
amount
$ \Delta E^{(\alpha)} = \epsilon\ \ql^{(0, \alpha)},$
where $\ql^{(0, \alpha)}$ is the ``link overlap'' between the states
``0'' and $\alpha$, defined by
\begin{equation}
\ql^{(0, \alpha)} = {1 \over N_b}\sum_{\langle i,j \rangle} S_i^{(0)}
S_j^{(0)}
S_i^{(\alpha)} S_j^{(\alpha)} ,
\end{equation}
in which the sum is over all the $N_b$ pairs where there are interactions.
Note that the {\em total} energy of the states is changed by an
amount of order unity. Fig.~\ref{e_vs_epsilon} shows the variation of the energy of
the ground state and that of an excited state which has an energy $\delta E$
above it in the absence of the
perturbation, as a function of
the strength of the perturbation. For the value
of $\epsilon$ shown the excited state has become the new ground state.

We denote the new ground state spin
configuration by $ \tilde{S}_i^{(0)}$, and indicate by
$\ql$,
with no indices, the link-overlap
between the new and old ground states,
i.e.
\begin{equation}
\ql = {1 \over N_b}\sum_{\langle i,j \rangle} S_i^{(0)}
S_j^{(0)}
\tilde{S}_i^{(0)} \tilde{S}_j^{(0)} .
\label{eq:ql}
\end{equation}
Similarly we denote the spin overlap by
$q$, where
\begin{equation}
q = {1 \over N}\sum_i S_i^{(0)}  \tilde{S}_i^{(0)}  ,
\label{eq:q}
\end{equation}
and the change in ground state energy by $\Delta E$, see
Fig.~\ref{e_vs_epsilon}.

From Fig.~(\ref{e_vs_epsilon}), we see that the change in ground state energy
satisfies the inequalities
\begin{equation}
\epsilon \ql \le \Delta E \le \epsilon .
\label{eq:delta_E_rels}
\end{equation}
Furthermore, the unperturbed
excitation energy $\delta E$ of the state which has become the
perturbed ground state (since
$\delta E+\epsilon q_l < \epsilon$,
see Fig.~\ref{e_vs_epsilon}) satisfies the conditions
\begin{equation}
0 < \delta E < \epsilon (1 - \ql) .
\label{eq:delta_e_rels}
\end{equation}

Note that
\begin{equation}
1 - q = {2 N_{\rm vol} \over N} ,
\label{eq:vol}
\end{equation}
where $N_{\rm vol}$ is the number of flipped spins, and
\begin{equation}
1 - \ql = {2 N_{\rm surf} \over N_b} ,
\label{eq:nsurf}
\end{equation}
where $N_{\rm surf}$ is the number of bonds crossed by the domain wall
bounding the flipped spins, i.e. the surface of the domain wall.

Next, following
Ref.~\onlinecite{palassini:00},
we discuss how quantities like $\ql$ and $q$ are expected to vary with
system size. Consider first those samples where a large excitation comprising
a finite fraction of the sites is generated, e.g. those where $\absq \le
0.5$. The average of $1-q$ over these samples given by
\begin{equation}
[1 - q ]^\prime\av \sim 1 ,
\label{eq:onemqp}
\end{equation}
where $[\cdots ]\av$ denotes an average over samples and the prime
denotes a restricted average over those samples with $\absq \le 0.5$.
From Eq.~(\ref{eq:nsurf}) we see
that the average of $1-\ql$ over these samples given by
\begin{equation}
[1 - \ql ]^\prime\av \sim L^{-\{d - d_s\} }.
\label{eq:onemqlp}
\end{equation}

Such a large excitation costs an energy of
order $L^{\theta'}$. This must be more than compensated for by
the energy
gained from the perturbation (relative to the ground state),
$\epsilon (1 - \ql)$, which, from Eq.~(\ref{eq:nsurf}),
is proportional to
$\epsilon L^{-(d-d_s)}$. Hence, assuming a
distribution of droplet energies with a finite weight at the origin,
the
probability that such a large excitation can be created is given by
\begin{equation}
P(\absq \le 0.5) \sim  L^{-\{ \theta' + (d - d_s)\} } .
\label{eq:frac}
\end{equation}

We now consider results of
averaging over {\em all}\/ samples rather than just those with
large droplets.
Eq.~(\ref{eq:frac}) shows that there is probability
$\sim  L^{-\{ \theta' + (d - d_s)\} }$ to have $1 - q$ of order unity,
and so,
assuming that samples with large droplets dominate the average, we have
\begin{equation}
[1 - q ]\av \sim L^{-\{\theta' + (d - d_s)\} }.
\label{eq:onemq}
\end{equation}
Similarly, we find
\begin{equation}
[1 - \ql ]\av \sim L^{-\mu_l },
\label{eq:onemql}
\end{equation}
where
\begin{equation}
\mu_l = \theta' + 2(d - d_s) .
\label{eq:mul}
\end{equation}

Note that that the ratio of the quantities in Eqs.~(\ref{eq:onemq}) and
(\ref{eq:onemql}) is independent of $\theta'$, i.e.
\begin{equation}
R \equiv {[1 - \ql ]\av \over [1 - q ]\av}
\sim L^{-\{d - d_s\} } .
\label{eq:r}
\end{equation}

Eq.~(\ref{eq:delta_e_rels}) shows that the droplet excitation energy
$\delta E$, see Fig.~\ref{e_vs_epsilon},
scales in the same way as $1 - \ql$, i.e. like
Eq.~(\ref{eq:onemqlp}) if only samples with $\absq \le 0.5$ are studied, and
like Eq.~(\ref{eq:onemql}) if all samples are averaged over.

In general we can extract $d-d_s$ from the surface to volume ratio of the
excited droplets of spins, but to get $\theta'$ we also need information about
the {\em probability}\/ that a large droplet is excited.

\section{Choice of Boundary Conditions}
\label{sec:bcs}

In those
two-dimensional models where the bonds form a ``planar graph'', which
unfortunately excludes periodic boundary conditions in both directions, very
efficient polynomial time ``matching'' algorithms can be
used\cite{bieche:80,barahona:82,palmer:99,middleton:01,hartmann:01a,hartmann:01}
to determine the ground state. As a result very large sizes can be
studied\cite{largest_sizes}.
By contrast, in three or higher dimensions the calculation is
known\cite{barahona:82} to be NP-hard which means that a polynomial-time
algorithm is unlikely to exist.

\begin{figure}
\includegraphics[width=\figurewidth]{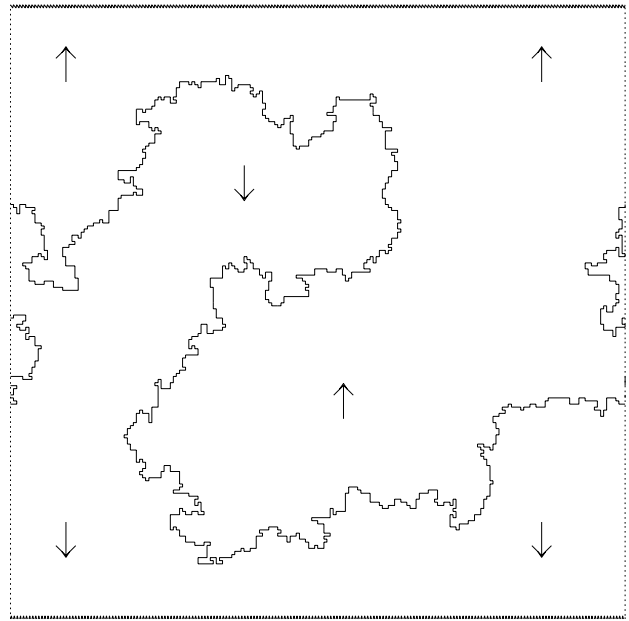}
\caption{
An excitation generated for an $L=200$ system obtained using the matching
algorithm. The vertical edges have periodic boundary conditions while the
horizontal edges have open boundary conditions.  With the matching algorithm
it is not possible to have periodic boundary conditions in both directions.
The line shows the boundary separating the region where the spins are flipped
when the perturbation is applied (denoted by ``$\downarrow$'')
from the region where they are not flipped (denoted by ``$\uparrow$'').
One sees that, in
this case, the excitation is a single
domain wall rather than a droplet.
}
\label{fig:samplewall}
\end{figure}

In Fig. \ref{fig:samplewall} we show the excitation generated, using the
matching algorithm,
for a $200 \times 200$
sample with
periodic boundary conditions on the vertical edges, and
open boundary conditions on the horizontal edges.
Unfortunately, the boundary of the
resulting excitation is not a droplet
but a single domain wall which
spans the whole system. This structure would not have been
possible if periodic
boundary conditions had also been applied along the top and bottom edges.
We find that a single domain wall is generated for a significant fraction of
the samples with open-periodic boundary conditions,
and so these boundary conditions are not appropriate for a study
of the {\em differences}\/ between droplets and domain walls\cite{moore;pc}.

One possibility to prevent the occurrence of domain walls 
is to first generate a ground state with open boundary conditions, and then
fix the orientations of the boundary spins\cite{middleton:01}
relative to each other by
introducing very strong bonds in favor of their relative
ground-state orientations. This still allows us to use the fast matching
algorithm and hence to treat large system sizes. We have only
tested this
approach briefly, since we found that
usually only very small droplets are
generated, with almost none spanning more than one quarter of the system
(i.e. with $\absq\le0.5$). The reason is that by fixing the boundary
spins an effective repulsion of the domain wall from the boundary is
introduced. This reduces the sizes of the droplets. Furthermore, 
it is also likely that this repulsion affects the morphology of
the domain walls.

We therefore decided to apply periodic boundary
conditions in both directions which,
unfortunately, means that the matching algorithm cannot be
used\cite{other_polynom}.
Instead we use
the spin glass ground state server at the University of
Cologne\cite{spin_glass_server}. This uses a ``branch and cut'' algorithm
for which, as we shall verify,
the CPU time increases faster than a power of the
system size, so we cannot study such large systems as are possible with
the matching algorithm\cite{largest_sizes}.
Nonetheless, the
implementation of the branch and cut algorithm on the Cologne server is very
efficient, so we can still study quite a large range of sizes, in practice $L
\le 64$.

\begin{figure}
\includegraphics[width=\figurewidth]{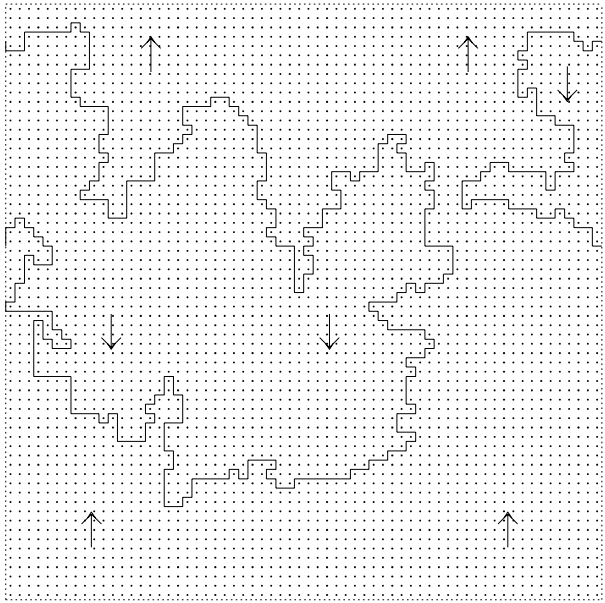}
\caption{
A droplet
excitation generated for an $L=64$ system obtained using the branch and cut
algorithm of the Cologne spin glass server.
Periodic boundary conditions are applied in both directions. The dots
indicate the lattice sites. 
Clearly the region where the spins are flipped forms a droplet.
}
\label{fig:wall_L64_1662461912}
\end{figure}

\begin{figure}
\includegraphics[width=\figurewidth]{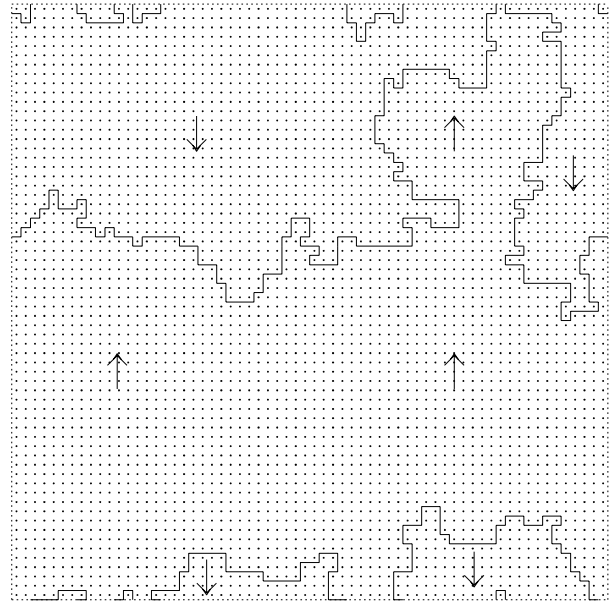}
\caption{
A ``sponge-like'' excitation
generated for an $L=64$ system with
periodic boundary conditions in both directions. 
}
\label{fig:wall_L64_818624358}
\end{figure}
 
Fig.~\ref{fig:wall_L64_1662461912} shows a large droplet
excitation created for an $L=64$ sample with periodic boundary conditions.
In some samples where large-scale excitations are produced, the region of
flipped spins wraps round the sample creating a ``sponge-like''
excitation\cite{houdayer:00b,houdayer:00}, which is, in effect, made up of
two domain walls\cite{sponge}.
An example is shown in
Fig.~\ref{fig:wall_L64_818624358}. Later we will discuss the role played
by sponges, finding that their properties are very similar to those of
large-scale, non-sponge excitations. 

\section{Results}
\label{sec:results}

\begin{table}
\begin{center}
\begin{tabular}{|r||r|r|r|r|r|r|r|r|r|}
\hline
$L$ & 4 & 6 & 8 & 12 & 16 & 24 & 32 & 44 & 64 \\
\hline
$N_{\rm samp}$ & 1973 & 1972 & 2059 & 1973 & 2098 & 1993 & 2099 & 2015 & 1763 \\
\hline
\end{tabular}
\end{center}
\caption{
The number of samples $N_{\rm samp}$ for each lattice size $L$.
}
\label{tab:samples}
\end{table}

All the numerical results are for a strength of
the perturbation $\epsilon = 2$. Asymptotically, the results should be
independent of the value of $\epsilon$. However, if we make $\epsilon$ too
large, then the asymptotic behavior will only be seen for large sizes, and if
$\epsilon$ is too small the statistics will be poor. Based on experience with
other models\cite{palassini:00},
the choice $\epsilon = 2$ seems to be a reasonable
compromise. 
In Table~\ref{tab:samples} we show the number of samples studied for each
lattice size.
\begin{figure}
\includegraphics[width=\figurewidth]{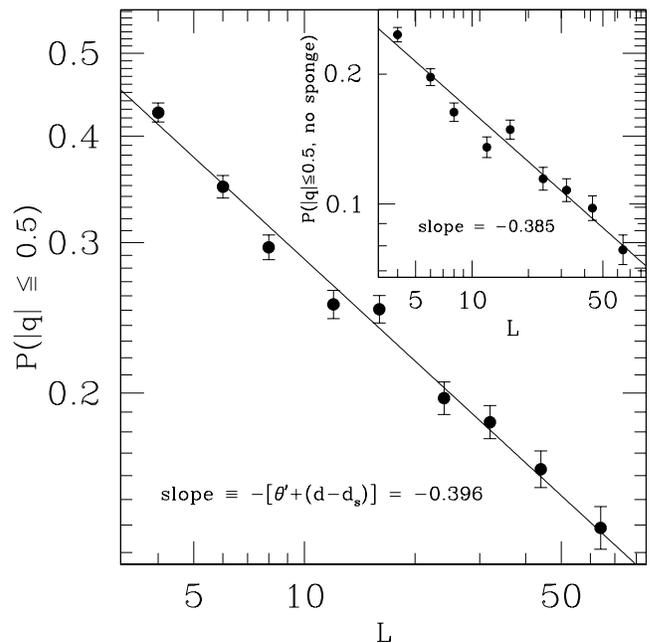}
\caption{
A log-log plot of the fraction of samples for which the spin overlap between the
perturbed and unperturbed ground states satisfies $\absq \le 0.5$. According to
Eq.~(\ref{eq:frac}) the slope should be $-[\theta' + (d -d_s)]$.
The inset is a log-log plot of the fraction of samples for which
not only is $\absq \le 0.5$ but also the excitation is
not sponge-like, see 
Fig.~(\ref{fig:wall_L64_818624358}). The slope is almost the same as that for
in the main part of the figure, indicating that a roughly fixed fraction,
about 55\%, of
excitations with $\absq \le 0.5$ are not sponges.
}
\label{frac}
\end{figure}

\begin{figure}
\includegraphics[width=\figurewidth]{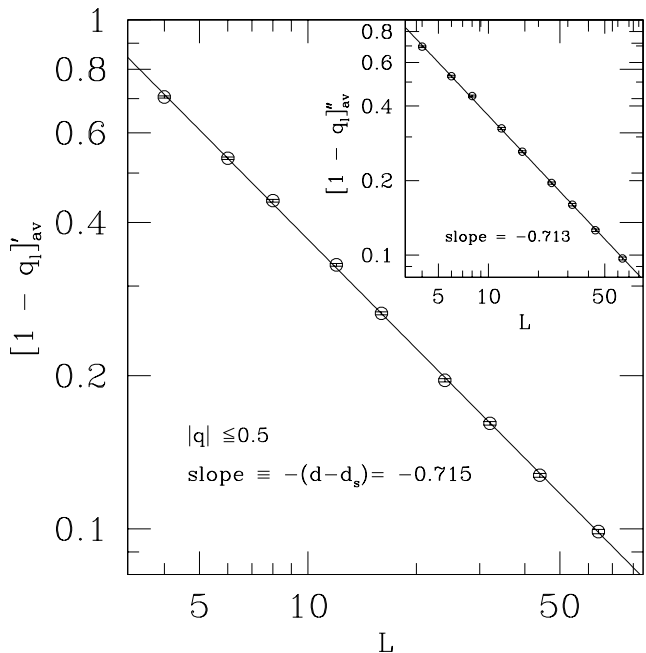}
\caption{
A log-log plot of the average of
$1 - \ql$ over those samples for which the spin overlap between
the perturbed and unperturbed ground states satisfy $\absq \le 0.5$. According
to Eq.~(\ref{eq:onemqlp}) the slope should be $-(d -d_s)$.
The inset shows results averaged over samples where not only is $\absq \le
0.5$ but also the excitation is not sponge-like, see
Fig.~\ref{fig:wall_L64_818624358}. The slope is virtually the same as that in
the main part of the figure, indicating that sponges have similar properties
to large non-sponge, droplet excitations. 
}
\label{qlp}
\end{figure}

\subsection{Average over samples with large droplets}
\label{subsec:large}
Firstly, we consider results obtained by averaging only over those samples
where large droplets were created, i.e. the spin overlap satisfies
$\absq \le 0.5$. Fig.~\ref{frac} shows results for the probability that
this occurs, while Fig.~\ref{qlp} shows the average of $1-\ql$ for
those samples. Fitting the data according to Eqs.~(\ref{eq:frac}) and
(\ref{eq:onemqlp}) we find
\begin{eqnarray}
d - d_s & = & 0.715 \pm 0.003 \nonumber  \\
\theta' + (d - d_s) & = & 0.396 \pm 0.015 ,
\label{eq:fits_prime}
\end{eqnarray}
which gives
\begin{eqnarray}
d_s & = & \ \ \, 1.285 \pm 0.003 \nonumber  \\
\theta' & = & -0.319 \pm 0.015 .
\end{eqnarray}
The value of $d_s$ agrees very well with best
estimates of around $1.27$, see Table~\ref{tab:exponents}, while
the value of $\theta'$ is a little more negative than the best
estimates for $\theta$ from domain wall calculations, 
which is about $-0.28$, see Table~\ref{tab:exponents}.
However, the difference is quite small, only about
two standard deviations. Hence we see that our results are consistent with the
droplet theory being correct in two dimensions. 

The insets to Figs.~\ref{frac} and \ref{qlp} show data for those samples where
the excitation is not only large, i.e. $\absq \le 0.5$, but is
also not
sponge-like, i.e. is {\em not}\/ of the form shown in
Fig.~\ref{fig:wall_L64_818624358}. The fraction of samples with $\absq \le
0.5$ for which the excitation is not sponge-like is roughly independent of
size at about 55\%. This is why the slope in the inset to Fig.~\ref{frac} is very
similar to that in the main part of the figure. Furthermore, the slope of the
data for the average of $1 - \ql$ in the inset to Fig.~\ref{qlp} is very
similar to the average over all samples with $\absq \le 0.5$ (shown in
the main part of
the figure), indicating that the properties of sponge-like excitation are
similar to those of large non-sponge droplets. 

We have also analyzed the data
including just the smaller sizes, $4 \le L \le 16$,
and just the larger sizes, $16 \le L \le 64$. and the results are shown in
Table~\ref{tab:fits_small_q}.
The values of $\theta'$ are, given the error bars, scarcely
more
negative than the best accepted value of $-0.28$.
One does not see a systematic trend over this range of sizes. In particular we
do not see the crossover, expected by Moore\cite{moore:02}, from a value of
about $-0.47$ at smaller sizes to about $-0.28$ at larger sizes.

\begin{table}
\begin{center}
\begin{tabular}{|r|c|c|c|c|}
\hline
&  $d_s$ & $\theta'$ & $Q_1$ & $Q_2$  \\
\hline
\hline
$4  \le L \le 64 $ & $1.285 \pm 0.003$ & $-0.319 \pm 0.015$ &
0.006 & 0.253 \\
$4  \le L \le 16 $ & $1.299 \pm 0.005$ & $-0.288 \pm 0.029$ &
0.037 & 0.050 \\
$16 \le L \le 64 $ & $1.285 \pm 0.008$ & $-0.298 \pm 0.043$ &
0.405 & 0.639 \\
\hline
\end{tabular}
\end{center}
\caption{
Fits for the data for those samples where $\absq\le0.5$ for different ranges of
sizes. $Q_1$ and $Q_2$ are the quality factors,
as conventionally defined\cite{press:95},
of the fits for
$[1 - \ql]'\av$ and $P(\absq\le 0.5)$ respectively. Note that $Q_1$ is
quite small for the fit with all sizes. This is because the error bars are
small and there are clearly some systematic effects. For example, if we omit
the L=4 point then $Q_1$ becomes $0.129$, which is much better. However, this
then gives $\theta =  -0.347 \pm 0.019$, which is further from the accepted
value of about $-0.28$ but agrees within the error bars with the value in the
table.
}
\label{tab:fits_small_q}
\end{table}

\begin{figure}
\includegraphics[width=\figurewidth]{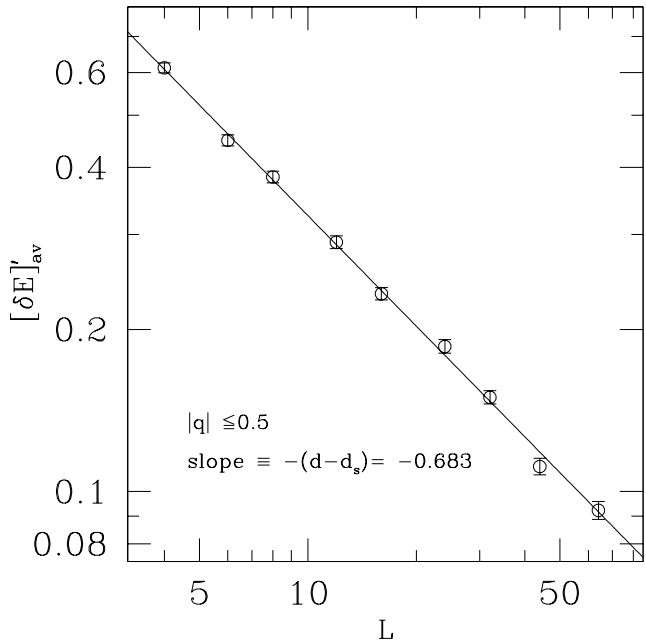}
\caption{
A log-log plot of the average of the unperturbed excitation energy,
$\delta E$, see Fig.~\ref{e_vs_epsilon},
over those samples for which the overlap between
the perturbed and unperturbed ground states satisfies $\absq \le 0.5$. According
to Eqs.~(\ref{eq:delta_e_rels}) and (\ref{eq:onemqlp})
the slope should be $-(d -d_s)$.
}
\label{delta_E}
\end{figure}

From Eq.~(\ref{eq:delta_e_rels})
we expect that the unperturbed excitation energy, $\delta E$, of the droplet
created by the perturbation, see Fig.~\ref{e_vs_epsilon},
should scale in the same way as $1 - \ql$, i.e. with an
exponent $d - d_s$ if restrict the average to those samples with $\absq \le
0.5$. The data, plotted in Fig.~\ref{delta_E}, gives $d - d_s =
0.683 \pm 0.011$ as compared with $0.715 \pm 0.003$ obtained from the data for
$1 - \ql$ in Fig.~\ref{qlp}. The difference, which is quite small, is
presumably due to corrections to scaling.

We should make clear why the average
of $\delta E$ induced by the perturbation
varies as $L^{-(d-d_s)}$ (if we only include
samples with $\absq \le 0.5$), whereas
the characteristic droplet energy varies as $L^{\theta'}$. The point is that
there is a {\em distribution}\/ of droplet energies which is assumed to have a
finite weight at the origin. Thus, some samples will have a large-scale
droplet with energy {\em much less}\/
than $L^{\theta'}$. In fact, the perturbed ground
state will only have a system-size droplet for those samples where the droplet
energy is comparable to the energy which can be gained from the perturbation,
i.e. $L^{-(d - d_s)}$.
The probability of this occurring is therefore $L^{-\{\theta' + (d - d_s)\}}$,
as shown in Fig.~\ref{frac}.

Our expectation in looking at samples with $\absq \le 0.5$ is that there is a
single large droplet which dominates, rather than having several smaller
droplets of comparable size, and we have verified that this indeed the case.
We have separated out the different droplets for each sample and looked at
averages over those {\em droplets}\/ containing $N/4$ spins or greater
(from Eq.~(\ref{eq:vol}) this corresponds to $\absq \le 0.5$). As an example,
for $L=64$, out of 1763 samples, 245 of them had $\absq \le 0.5$ and 241 had a
single droplet which would give $\absq \le 0.5$ by itself.
If we repeat the finite-size scaling analysis for these large
droplets we find $d -d_s = -0.715 \pm 0.003$ and $\theta' + (d - d_s) =
-0.392 \pm 0.015$, in very close agreement with the results in
Eq.~(\ref{eq:fits_prime}).


\begin{figure}
\includegraphics[width=\figurewidth]{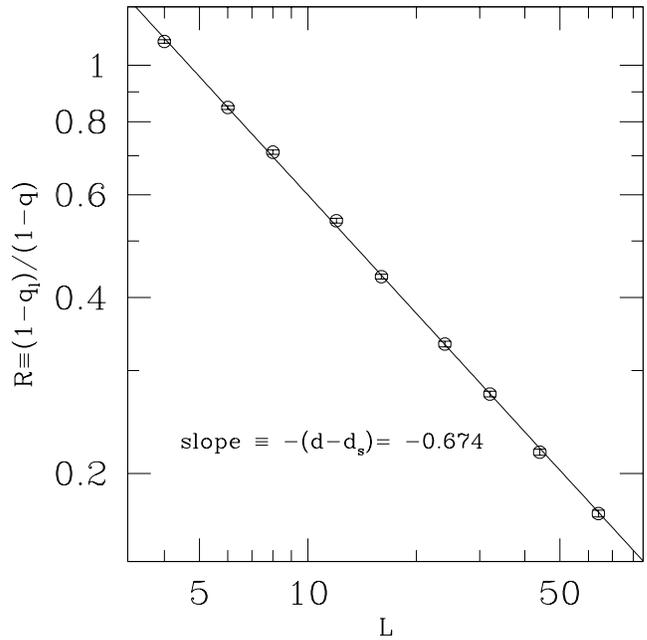}
\caption{
A  log-log plot of the ratio $[1 - \ql]\av/ [1 - q]\av$ against $L$
obtained by
averaging over {\em all}\/ samples. From Eq.~(\ref{eq:r}) the slope should be
$-(d - d_s)$.
}
\label{r}
\end{figure}

\begin{figure}
\includegraphics[width=\figurewidth]{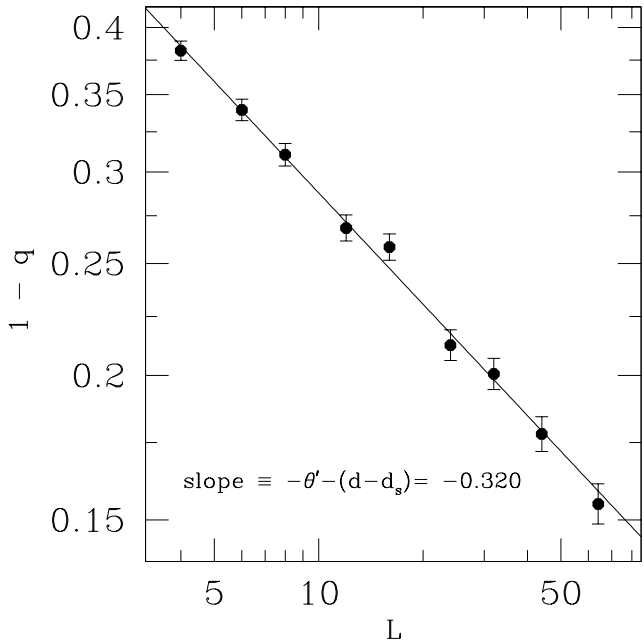}
\caption{
A log-log plot of $ [1 - q]\av$ against $L$ obtained by
averaging over {\em all}\/ samples. From Eq.~(\ref{eq:onemq})
the slope should be
$-[\theta' + (d - d_s)]$.
}
\label{onemq}
\end{figure}

\subsection{Average over all samples}
\label{subsec:all}

Next we discuss the results obtained by averaging over {\em all}\/ samples,
rather than just those with $\absq \le 0.5$. Fig.~\ref{r} shows results for the
ratio of the average of $1 - \ql$ to the average of $1 - q$, and
Fig.~\ref{onemq} shows results for the average of $1-q$. Fitting the data
according to Eqs.~(\ref{eq:r}) and (\ref{eq:onemq}) we find
\begin{eqnarray}
d - d_s & = & 0.674 \pm 0.004 \nonumber \\
\theta' + (d - d_s) & = & 0.320 \pm 0.010 ,
\end{eqnarray}
which gives
\begin{eqnarray}
d_s & = & \ \ \, 1.326 \pm 0.004 \nonumber  \\
\theta' & = & -0.354 \pm 0.010 .
\end{eqnarray}

Analyzing the data for different ranges of sizes gives the results shown in
Table~\ref{tab:fits_all_samps}. We note that the values for $d_s$ are
larger than those obtained from analyzing just large droplets, see
Table~\ref{tab:fits_small_q}, and the values for $\theta'$ somewhat more
negative. This is presumably
due to the effects of averaging over droplets of
different sizes, as discussed by Middleton\cite{middleton:01}.
Indeed, our results for $\mu_l \equiv 2(d-d_s) + \theta'$, see
Table~\ref{tab:fits_all_samps}, agree quite well with
those for smaller sizes in Fig.~4 of Ref.~\onlinecite{middleton:01}, even
though our boundary conditions are different from his.
Nonetheless, the
variation of exponents with range of $L$ is too small to yield
$\theta'$ about $-0.47$ as
suggested by Moore\cite{moore:02},
thought the trend ($\theta'$ more negative for smaller
lattice sizes and when we average over different droplet sizes) is in the
right direction.

From Table~\ref{tab:fits_all_samps}, we find $\theta' =
-0.364 \pm 0.021$ when averaging over all samples and just
including the smaller
sizes. This is about $-0.08 \pm 0.02$ different from the best value of
$-0.28$, see Table~\ref{tab:exponents}.
Furthermore, a reanalysis\cite{palassini:02} of the data of
Ref.~\onlinecite{palassini:00}, just keeping those samples with small $\absq$,
i.e. similar to our analysis which yielded the fits in
Table~\ref{tab:fits_small_q}, found the same discrepancy in three-dimensions
as the original analysis\cite{palassini:00}
which averaged over droplets of all sizes. We see
that the discrepancy in the value for $\theta'$ in
Table~\ref{tab:fits_small_q}, relative to $-0.28$, is much smaller than that in
Table~\ref{tab:fits_all_samps}, even if the fits are restricted to smaller
sizes.
Hence a much larger effect would be needed in three dimensions to
explain why the numerics found $\theta' \simeq 0$ if the correct result is
$\theta' = \theta \simeq 0.20$. In four-dimensions there is an even greater
discrepancy\cite{palassini:00}, $\theta' \simeq 0, \theta \simeq 0.70$.

\begin{table}
\begin{center}
\begin{tabular}{|r|c|c|c|}
\hline
&  $d_s$ & $\theta'$ & $\mu_l$ \\
\hline
\hline
$4  \le L \le 64 $ & $1.326 \pm 0.004$ & $-0.354 \pm 0.010$ &
$0.995 \pm 0.008$ \\
$4  \le L \le 16 $ & $1.340 \pm 0.007$ & $-0.364 \pm 0.021$ &
$0.956 \pm 0.017$ \\
$16 \le L \le 64 $ & $1.322 \pm 0.011$ & $-0.317 \pm 0.030$ &
$1.039 \pm 0.024$\\
\hline
\end{tabular}
\end{center}
\caption{
Fits for the data obtained by averaging over all samples
for different ranges of
sizes. Here $\mu_l \equiv 2(d - d_s) + \theta'$.
}
\label{tab:fits_all_samps}
\end{table}

\begin{figure}
\includegraphics[width=\figurewidth]{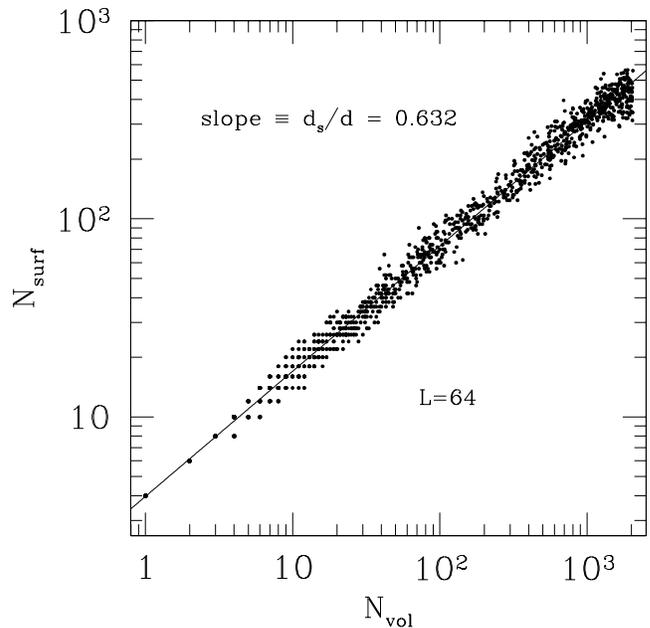}
\caption{
A plot of the size $N_{\rm vol}$ and surface area $N_{\rm surf}$ of
individual droplets for $L=64$. A fit gives $d_s/d = 0.632$.
}
\label{q_ql_L64}
\end{figure}

To get an independent estimate of $d_s$ we have also looked at the size,
$N_{\rm vol}$, and surface area, $N_{\rm surf}$, of individual droplets for a
given size. Fig.~\ref{q_ql_L64} shows data for the largest size $L=64$, and
similar results are found for other sizes. Doing a least squares fit
gives $d_s/d = 0.632$, or $d_s = 1.264$, in very good agreement with the best
estimates of
around $1.27$, see Table~\ref{tab:exponents}.

\subsection{Computational Complexity}
\label{subsec:cputime}

\begin{figure}
\includegraphics[width=\figurewidth]{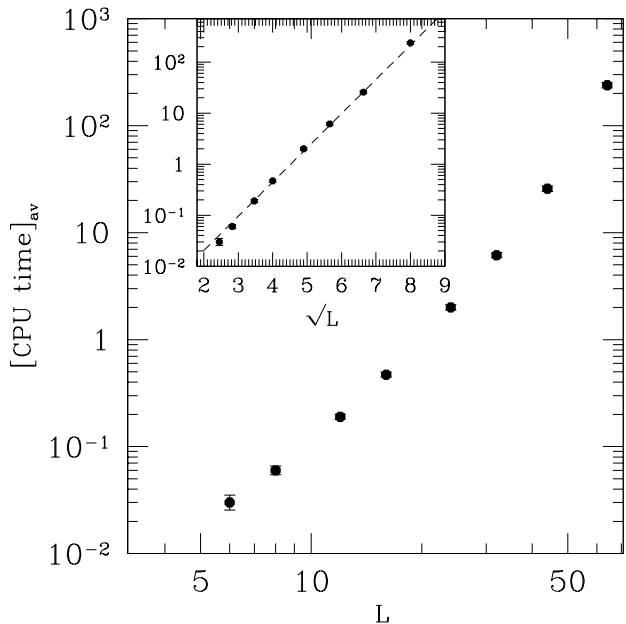}
\caption{
A log-log plot of the 
average CPU time per sample as a function of the system size
$L$.
The upwards curvature shows that the CPU time increases faster than a power of
$L$ as expected.  The inset shows that the CPU varies roughly as
$\exp ({\rm const.}\, \sqrt{L})$ for the larger sizes. The dashed line is a
guide to the eye.
}
\label{cputime}
\end{figure}

Finally we discuss how the CPU time depends on the system size.
Fig.~\ref{cputime}  shows a log-log
plot of the average
CPU-time on the Sun Ultra 10/440
of the Cologne ground state
server against $L$.
The clearly visible upwards curvature indicates that the time increases faster
than a power of $L$, i.e. the algorithm is non-polynomial, as expected.
The inset to
the figure shows that, empirically, the CPU time varies like $\exp ({\rm
const.}\, \sqrt{L})$ for the larger sizes studied. However, we do not know the
true asymptotic dependence of the CPU time on system size.

\section{Conclusions}
\label{sec:conclusions}

We have studied large-scale, low-energy excitations at $T=0$ in the Ising spin
glass in two-dimensions with periodic boundary conditions.
We have particularly looked at the dependence of
$d_s$ and $\theta'$ on whether or not we average over droplets of different
sizes, and on the range of system sizes that are included in the fits.
Averaging just over those samples in which large droplets are generated, see
Sec.~\ref{subsec:large} and Table~\ref{tab:fits_small_q},
the results do not vary much if we consider
different ranges of
system sizes, and are consistent with the droplet theory prediction that
$\theta' = \theta$. If we average over all samples,
see Sec.~\ref{subsec:all} and Table~\ref{tab:fits_all_samps},
which, for each lattice
size, involves averaging over a range of droplet sizes, the results vary more
as a function of
the range of system size considered. However, the effect is too small to give
an effective value of $\theta'$ close to $-0.47$ as has been recently
proposed\cite{moore:02}, though our result in line 2 of
Table~\ref{tab:fits_all_samps} does lie between $-0.47$ and the
expected value of about $-0.28$. It is possible that different ways of
generating droplets will yield different effective values of $\theta'$ for
small sizes, and this may be the
reason why Refs.~\onlinecite{liang:92,kawashima:92a,kawashima:92}
find $\theta' \simeq -0.47$.

We have also considered the role of sponge-like excitations, see 
Fig.~\ref{fig:wall_L64_818624358},
finding that their properties are very similar to those of large,
non-sponge droplets, see Fig.~\ref{fig:wall_L64_1662461912}.

Overall, our results
support the droplet theory for the two-dimensional spin glass and
do not provide a straightforward way to reconcile the numerical
results\cite{krzakala:00,palassini:00} used to support the TNT scenario
with the prediction of the droplet theory that
the stiffness exponent is asymptotically
the same for droplets and domain walls.
For this prediction to be true,
larger corrections to finite-size scaling would be needed in
three and four dimensions than in two dimensions.

\begin{acknowledgments}
We would like to thank M.~A.~Moore for a fruitful correspondence and for
comments on an earlier version of the manuscript. We would also like to thank
A.~A.~Middleton, M.~Palassini, D.~A.~Huse and F.~Ritort
for comments on earlier drafts. APY acknowledges support
from the National Science Foundation under grant DMR 0086287 and the EPSRC
under grant GR/R37869/01. He would also like to thank David Sherrington for
hospitality during his stay at Oxford. AKH obtained
financial support from the DFG (Deutsche Forschungsgemeinschaft) under grants
Ha 3169/1-1 and Zi 209/6-1. We are extremely grateful to the group of Michael
J\"unger at the University of Cologne for putting their spin glass ground
state server in the public domain, and for not complaining even when we used a
substantial amount of computer time on it. We would particularly like to thank
Thomas Lange and Frauke Liers for helpful correspondence on the use of the
server. Also computer resources from the Paderborn Center for Parallel
Computing (Germany) and the Institut f\"ur Theoretische Physik of the
Universit\"at Magdeburg (Germany) were used.

\end{acknowledgments}

\bibliography{refs,comments}

\end{document}